# The role of individual characteristics of human subjects on the radiation burden of the bronchial airways from radon progeny

Péter Füri[1,*], Árpád Farkas[1], Werner Hofmann[2], Balázs G. Madas[1]

[1] Environmental Physics Department, Institute for Energy Security and Environmental Safety, HUN-REN Centre for Energy Research, Budapest, Hungary

[2] Department of Chemistry and Physics, Paris Lodron University, Salzburg, Austria

**Corresponding author:**

*Péter Füri

E-mail address: peter.furi@ek.hun-ren.hu

Tel. +36-1-392-2222 (ext. 3254)

**Abstract.** Variability in radiation-related health risk and genetic susceptibility to radiation effects within a population is a key issue for radiation protection. Besides differences in the health and biological effects of the same radiation dose, individual variability may also affect dose distribution and its consequences for the same exposure. As exposure to radon progeny affects a large population and has a well-established dose–effect relationship, investigating individual variability upon radon exposure may be particularly important. Using the Stochastic Lung Model combined with mucociliary clearance and alpha-particle microdosimetry models, deposition rates and absorbed dose rates were determined for a healthy adult, a 5-year-old child, and an adult with severe asthma. The results show that children receive significantly higher absorbed dose rates in basal and secretory cell nuclei than healthy adults, despite lower deposition rates, due to smaller airways and thinner mucus layers. For individuals with severe asthma, both deposition rates and dose rates are higher due to airway contraction and slower mucus clearance, although increased mucus thickness reduces absorbed dose rates. These findings demonstrate that anatomical and physiological differences significantly influence absorbed doses in the lungs upon radon exposure and highlight the importance of accounting for individual variability in radiation protection and risk assessment.

Short title: Individual differences in lung dosimetry upon radon exposure

**Acknowledgments**

This study is part of a project that has received funding from the Euratom research and training programme 2019-2020 under grant agreement no 900009 (RadoNorm).

# 1. Introduction

Variability in radiation-related health risk and genetic susceptibility to radiation effects within a population is an important issue for radiation protection. Differences in radiation sensitivity, if significant, raise the ethical and policy question as to whether some individuals or groups are inadequately protected by the present system and regulations, and whether it would be acceptable to apply different exposure limits for various subgroups of the population or ultimately at the individual level (Kreuzer et al. 2018; Bouffler et al. 2019).

Besides the differences in the health and biological effects of the same radiation dose, individual variability may also affect the dose consequences of the same exposure, which is mostly considered in medical exposures (e.g., Deak et al. 2010). This latter aspect can be particularly important in case of internal exposures where the biokinetics of incorporated radionuclides – a major determinant of the dose distribution – is also affected by anatomical and physiological differences (Kreuzer et al. 2018; Bouffler et al. 2019). Most of the research efforts on individual sensitivity in radiation protection, however, are focusing on variability in the effects of the same dose.

As there are large uncertainties in the long-term effects of low dose radiation even if individual differences are not considered, studies on individual sensitivity have to focus on such exposure scenarios, (i) where a large population is affected, and (ii) there is a well-established dose effect relationship for the general population. Exposure to radon progeny meets both requirements as it contributes to about 50% of the natural background radiation, and it is second most important cause of lung cancer (NRC, 1999) with an established dose effect relationship down to 100 Bq/m$^3$ (Darby et al. 2005, 2006).

The objective of the present study is to compare the dose distributions in different individuals upon exposure to radon progeny. While differences between individuals include sex, age at exposure, state of health, genetic and epigenetic make-up, lifestyle, and attained age, here we focus on two aspects only: age and a specific lung disease, i.e. severe asthma. Deposition rates and absorbed dose rates upon radon exposure are determined for three subjects with different breathing patterns and airway geometry. The first person represents a healthy adult Caucasian man with height of 176 cm and a tidal volume of 500 cm$^3$. The second subject is a 5-year-old child with a height of 110 cm, and a tidal volume of 213 cm$^3$ (ICRP, 1994). The third subject is an adult Caucasian man with severe asthma with contracted airways and increased mucus layer thickness.

# 2. Methods

*2.1 The particle deposition model and its input data*

The Stochastic Lung Model or IDEAL 2.0-code (SLM in the followings, Hofmann and Koblinger 1990, 1992; Koblinger and Hofmann 1990) was applied to determine the deposition distribution of inhaled radon progeny at regional and airway generation level. This model was successfully validated (Hofmann 2011) and used in the past to determine the deposition distribution of airborne bacteria (Balásházy et al. 2009), aerosol medicines (Farkas et al. 2016), radon progeny (Füri et al. 2020a), ultrafine urban particles (Füri et al. 2020b) and Severe Acute Respiratory Syndrome Coronavirus 2 (SARS-CoV-2, Madas et al. 2020) in the human airways.

In the SLM, the structure of the bronchial airways is generated by Monte Carlo methods using the anatomical dataset from Raabe et al. (1976). The acinar region is built according to the description and data from Haefeli-Bleuer and Weibel (1988). The model is able to provide deposition distribution of inhaled particles over airway generations defined by the number of bifurcations from the trachea. Due to the applied Monte Carlo method, it can describe both the intra- and intersubject variabilities of the human airway structure.

To simulate paediatric airways, the lengths and diameters of the airways have to be scaled down. For this purpose, the approach by Phalen and Oldham (2001) was used assuming that a typical adult male is characterized by a height of 176 cm, while the same quantity is 110 cm for a 5-year-old child (ICRP Publication 66, 1994).

Deposition fraction in an airway segment (region or airway generation) was calculated as the ratio of the number of progeny deposited in that segment to the number inhaled progeny. By the same token, deposition rate was defined as the product of the number of inhaled particles per hour and the deposition fraction. Deposition calculations were performed by simulating the effects of Brownian diffusion, impaction, and gravitational settling. The probability of extrathoracic deposition in the impaction regime was determined by the formula of Yu (1981), while in the diffusion regime by the formula of Cheng et al. (1996). In order to obtain good enough statistical power, the pathway of ten thousand particles were simulated in each run.

To realistically simulate the deposition of radon progeny in diseased lungs, both the specific breathing mode and the contraction of the bronchial airways have to be taken into account. For this purpose, an asthma model was integrated into a previous version of the Stochastic Lung Model (Füri et al. 2017). The extended model considers the modified airway geometry and the modified air velocity values according to the degree of disease severity. The probability and the extent of airway contraction (relative reduction of airway diameter) was assessed based on spirometric data of patients with asthma. These probability contraction and extent of contraction values are listed in Table 1.

Table 1. The probability and extent of the contraction of the airways in the bronchial region

| Airway generation | Probability of contraction | Extent of contraction |
| --- | --- | --- |
| 1$^{st}$ | 0 % | 0 % |
| 2$^{nd}$ | 10 % | 3 % |
| 3$^{rd}$ | 20 % | 5 % |
| 4$^{th}$ | 30 % | 8 % |
| 5$^{th}$ | 40 % | 10 % |
| 6$^{th}$ | 50 % | 15 % |
| 7$^{th}$ | 60 % | 20 % |
| 8$^{th}$ | 75 % | 20 % |
| from 9$^{th}$ to 21$^{st}$ | 75 % | 25 % |

*2.2 Breathing parameters*

All the parameters characterizing the breathing pattern for healthy subjects were taken from ICRP Publication 66 (1994) except the tidal volume which was taken from Pleil et al. (Pleil et al. 2021). Only nose breathing was considered. The tidal volume at sitting is usually not altered significantly for people with asthma. The functional residual capacity (FRC) and the breathing frequency, however, is higher for asthmatic patients. The values can be found in Table 2.

Another characteristic of severe asthma is that the breathing pattern is asymmetric with a shorter inhalation and longer exhalation time. Therefore, the duration of inhalation and exhalation was supposed to be 1 s and 1.4 s, respectively, for asthmatic patients, as opposed to the 2,5-s-long duration of for the healthy adult, and 1.2-s-long durations for five-years-old children. The breathing parameters are also summarized in Table 2.

Table 2. Breathing parameter values used in the present work corresponding to an adult man, a 5-year-old child and an adult asthmatic patient while sitting. FRC – functional residual capacity; VT – tidal volume; $f_B$ – breathing frequency

| Subject | FRC [$cm^3$] | VT [$cm^3$] | $f_B$ [$min^{-1}$] | Inhaled air volume [$m^3/h^{-1}$] |
|---|---|---|---|---|
| adult man | 3300 | 500 | 12 | 0.36 |
| 5-year-old child | 767 | 213 | 25 | 0.32 |
| asthmatic patient | 4000 | 500 | 25 | 0.75 |

*2.3 The size- and activity-distribution of the radon progeny*

The simulations were performed for a dwelling with a radon activity concentration of 40 Bq/$m^3$. The activity concentration ratio of $^{218}$Po, $^{214}$Pb, and $^{214}$Bi isotopes was supposed to be 0.58/0.44/0.29 (UNSCEAR, 2000) corresponding to an equilibrium factor of 0.4. Unattached progeny are responsible for 6% of the total potential alpha energy concentration (PAEC, Haninger 1998). 90% of unattached PAEC is linked to the decay of $^{218}$Po isotopes and the remaining 10% to that of $^{214}$Pb (ICRU, 2012). The activity median aerodynamic diameter (AMAD) of attached progeny was supposed to be 230 nm, while the activity thermodynamic diameter (AMTD) of unattached progeny was supposed to be 0.8 nm (Marsh et al. 2005).

*2.4 The particle clearance and the dosimetry model*

In order to estimate absorbed doses in different airway generations, mucociliary clearance of deposited radon progeny and radiation transport of alpha particles emitted during the decays also have to be considered. Descriptions of these models are presented below.

After deposition, radon progeny immediately starts to move with the mucus layer. The $^{218}$Po, $^{214}$Pb, and $^{214}$Bi progeny are tracked until they leave the trachea or decay into $^{210}$Pb. The airway generations where alpha decays (from $^{218}$Po and $^{214}$Po) take place are recorded for each deposited particle. The mucus velocity in the trachea was supposed to be 11.8 mm/min for healthy adults and 6.3 mm/min for asthmatic patients (Mezey et al. 1978), while 2.7 mm/min for the 5-year-old child (Sturm 2012). The mucus velocity in all subsequent airway generations decreases by a factor of 0.67 compared to the previous airway generation (Hofmann and Sturm 2004).

The thickness of the mucus layer in the 3$^{rd}$ airway generation of healthy adults was supposed to be 5 µm atop of a 6-µm-thick sol layer with cilia (ICRP, 1994). For patients with severe asthma, the mucus layer is usually much thicker due to mucus hypersecretion and airway contraction (Rogers 2004). In the present study, a three-times thicker mucus layer is supposed relative to healthy adults. For 5-year-old children, the thicknesses of the mucus and cilia layers were scaled down by the height ratio.

The location of the target cells was selected randomly in the airway epithelium in 25 different depths ranging from 2 to 50 μm (measured from the top of the epithelium). The depth distribution of the basal or secretory cells was determined according to Mercer et al. (1991). They distinguished large bronchi (with diameters larger than 3 mm), bronchi (with diameters smaller than 3 mm but larger than 1 mm), and terminal bronchioles (with diameters smaller than 1 mm). The depth distribution of the basal and secretory cells is presented in Figure 1.

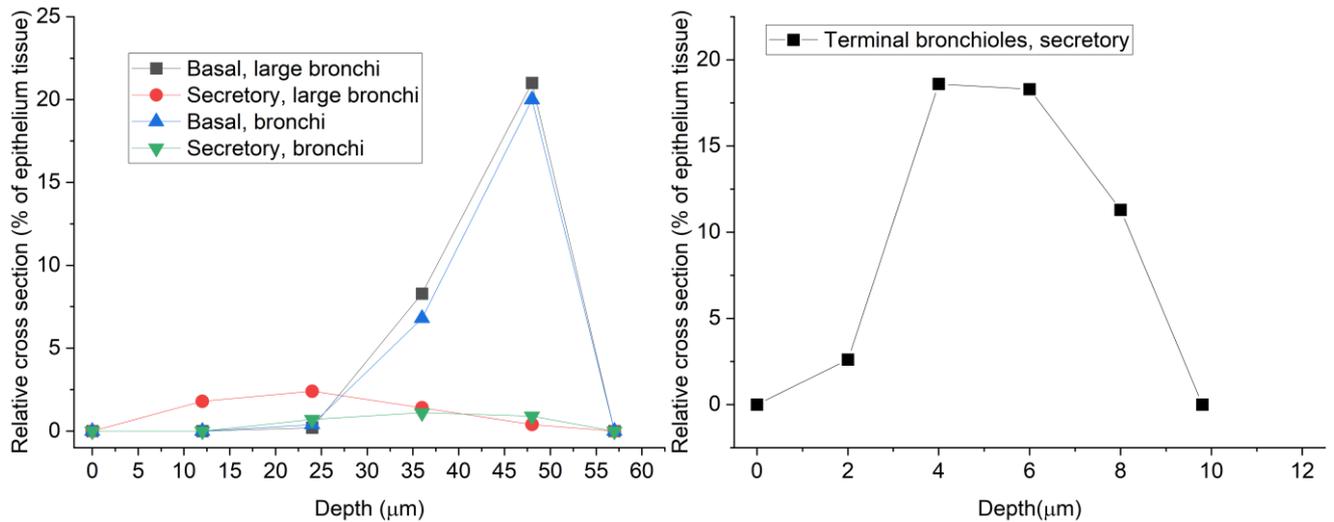

Figure 1. The depth distribution of basal and secretory cells in large bronchi, bronchi, and terminal bronchioles. There are no basal cells in the terminal bronchioles.

Absorbed doses in cell nuclei were calculated based on the locations of alpha decays. First, the average energy absorbed during one hit in a single target cell nucleus (basal or secretory) was determined in all the bronchial airway generations for both decay energies, 6 MeV and 7.69 MeV. $10^5$ progeny and target cells were placed in each airway generation. Considering a cylindrical coordinate system, the first coordinate (radius) of the location of $^{218}$Po and $^{214}$Po decays were selected randomly from a uniform distribution over a layer representing the mucus, while the second (azimuth angle) and third coordinate (height) are selected randomly from a uniform distribution over the surface of the airway. In order to spare computation time, energy absorbed in target cells were calculated with restriction that the direction of the movement of alpha-particles results in a target cell hit.

The probability that an alpha particle reaches a target cell nucleus was determined by the method described by Crawford-Brown and Shyr (1987). These probabilities were than used as weighting factors to determine absorbed doses in cell nuclei. Finally, the average absorbed energy in the nuclei of the basal and secretory cells was multiplied with the total number of alpha-decays of the $^{218}$Po and $^{214}$Po isotopes.

## 3. Results and discussion

### 3.1 Computed deposition fractions and deposition rates

The deposition fraction of attached progeny in the extrathoracic region is rather low: 7.89% for healthy adult men, 8.32% for 5-year-old children, and 6.23% for patient with severe asthma. In contrast, unattached progeny deposit with high probability in this region of the respiratory tract: the deposition fraction is 95.9% for healthy adult men, 96.8% for 5-year-old children, and 91.88% for asthmatic patients. These numbers show that the majority of unattached progeny deposit in the upper airways. The decrease of extrathoracic deposition of unattached progeny can be explained by the lower deposition by diffusion and sedimentation. Higher flow rate during inhalation shortens the time available for particles to deposit in the upper airways and in the lungs. Although deposition via impaction is increased by higher air velocities, it does not compensate for the effects of the other two mechanisms.

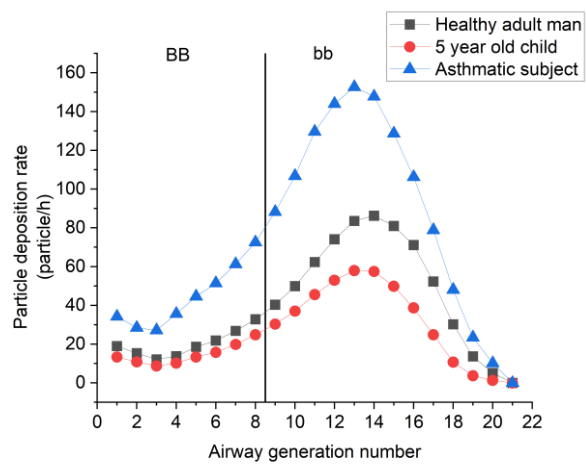

Figure 2. Deposition rates of radon progeny (both attached and unattached) in the bronchial airways of healthy adults, 5-year-old children and adults with severe asthma.

Figure 2 shows the distribution of deposition rates over the airway generations. The highest deposition rate can be found in the $14^{th}$ airway generation for healthy adults and in the $13^{nd}$ airway generation for children and asthmatic subjects. It also shows that the deposited number of radon progeny is higher for a healthy adult man than for a 5-year-old child in all the bronchial airways. While the volume of inhaled air is similar (0.36 $m^3$/h for adults vs. 0.32 $m^3$/h for children), the breathing frequency is more than doubled in children (25/min vs. 12/min) decreasing the effectiveness of diffusion and sedimentation and so resulting in lower deposition fractions.

Figure 2 also shows that deposition rates are much higher for asthmatic people than for healthy adults. There are two reasons behind the difference. In addition to the higher number of inhaled radon progeny in asthmatic patients, deposition fractions in most of the bronchial airways are also higher in people with severe asthma. The latter is related to the contraction of the airways which increases the effectiveness of both diffusion and sedimentation.

### 3.2 Absorbed dose rates in basal and secretory cell nuclei

To highlight the effect of individual characteristics on the radiation burden of the large bronchial airways, absorbed dose rates in the nuclei of basal and secretory cells of the bronchial airways in a healthy adult man, a 5-year-old child, and an adult with severe asthma are presented in case of the same exposition, i.e. a one-hour-long exposure in radon concentration of 40 Bq/m$^3$ with an equilibrium factor of 0.4.

The major determinant of absorbed doses is the number of alpha decays in the investigated airway. It depends on the deposition rate of radon progeny, and the time available for decay in the given airway. The latter can be calculated by the mucus velocity and the length of the airway. It is important to note that $^{214}$Po (with a 7.69 MeV alpha-particle) has a half-life of 0.164 microseconds which is negligible compared to the time spent in any airway. However, all the half-lifes of $^{218}$Po (with a 6.00 MeV alpha-particle), $^{214}$Pb and $^{214}$Bi have to be considered to calculate the amount of $^{214}$Po generated in each airway.

The mucus thickness and the airway diameter are other important factors. Alpha particles have to cross the airway lumen and the mucus to reach the airway epithelium. Increasing the diameter of the airway and thickness of the mucus layer both lead to a longer intersection of the alpha-track with material where energy absorbed does not result in absorbed dose in the target cell nuclei. If the intersection is too long, the alpha particle loses all their energy before reaching any target cell.

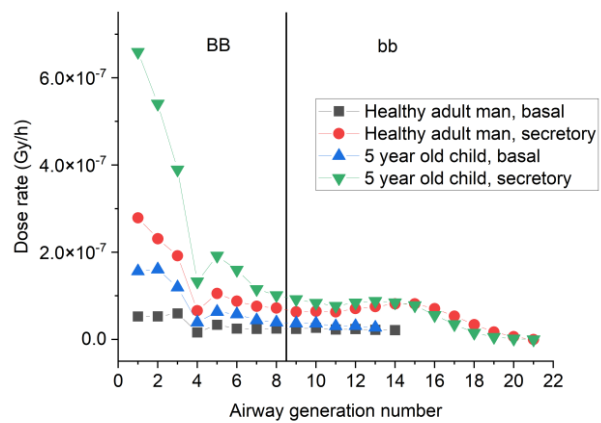

Figure 3. Absorbed dose rates in basal and secretory cell nuclei in different airway generations of a 5-year-old child and a healthy adult man

Figure 3 shows absorbed dose rates in basal and secretory cell nuclei in different airway generations of healthy adults and 5-year-old children. The radiation burden of the same exposure is much higher in the large bronchial airways (BB) of a 5-year-old child, than a healthy adult. This is the result of the smaller airways and the thinner mucus thickness in children compared to adults. These results show that the difference in age linked to a difference in height and breathing pattern, strongly affects the radiation burden of the bronchial epithelium.

From the 16$^{th}$ airway generation in the small bronchi (bb), however, absorbed dose rates are higher in adults (in case of secretory cells). While the hit probability and the average amount of absorbed energy for one alpha-particle is higher for children as in case of the BB, the particles inhaled are not transported so deep into the lung due to the lower tidal volume, and therefore the number of radon progeny deposited in the bb and in acinar region is much lower for 5-year-old children, than for healthy adults.

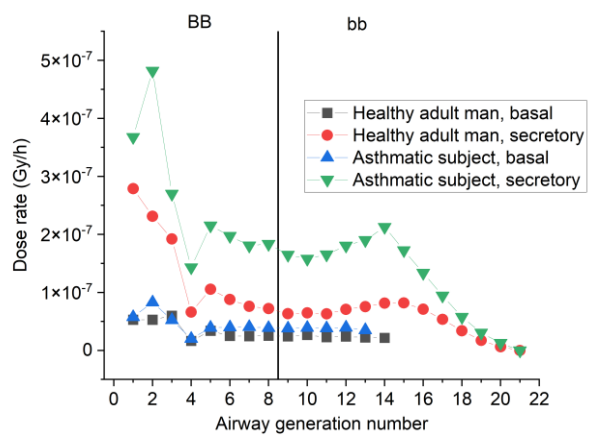

Figure 4. Absorbed dose rates in basal and secretory cell nuclei in different airway generations of a healthy adult man and a man with severe asthma

Figure 4 shows absorbed dose rates in basal and secretory cell nuclei in different airway generations of healthy adults and adults with severe asthma. In most cases, asthma results in significantly higher dose rates. Besides the higher deposition rates, hit probabilities and average absorbed energies in target cell nuclei from a single alpha particle are also higher due to contracted state of the airways. In addition, the slower movement of mucus within airways of the same length also allows more time for decay in asthmatic airways. These all increases absorbed dose rates.

The increased mucus layer thickness, however, decreases absorbed dose rates in people with severe asthma, as it reduces both hit probabilities and average absorbed energies from a single alpha-particle. This kind of shielding effect, however, does not compensate for the dose enhancing effects of other differences. If the mucus in asthmatic patients is thinner than it was supposed, the shielding effect is decreasing, so the difference between healthy and asthmatic adults is even larger, than what can be seen in Figure 4.

## 4. Conclusions

The comparison of absorbed doses in target cell nuclei of children, healthy adults, and adults with severe asthma clearly shows that age and lung diseases also affect absorbed doses from the same exposure. Inhaling the same air with a radon activity concentration 40 Bq/m$^3$, results in different deposition rates of radon progeny and different dose rates in the bronchial airways from their decay. It shows that individual sensitivity to radiation exposure not only manifests itself in differences in the health and biological effects of the same radiation dose, but also in the dose consequences of the same exposure. In this way, the individual risk upon a given exposure not only depend on the cellular sensitivity to ionizing radiation, but also on the anatomy and physiology of the individual modulating absorbed doses. It is particularly relevant in case of internal exposures like exposure to radon progeny.

**Disclosure of interests**

The authors report no conflicts of interest.

## Contributions

ÁF, WH and BGM conceived and designed the study, PF implemented the models (SLM, mucociliary clearance, & microdosimetry) applied in the study, performed the simulations, and visualized the data, ÁF, PF, WH, and BGM analyzed and interpreted the results and wrote the manuscript.

## References

Balásházy I, Horváth A, Sárkány Z, et al (2009) Simulation and minimisation of the airway deposition of airborne bacteria. Inhal Toxicol 21:1021–1029. https://doi.org/10.1080/08958370902736646

Bouffler S, Auvinen A, Cardis E, et al (2019) Strategic Research Agenda of the Multidisciplinary European Low Dose Initiative (MELODI) – 2019. Multidisciplinary European Low Dose Initiative

Cheng K-H, Cheng Y-S, Yeh H-C, et al (1996) In vivo measurements of nasal airway dimensions and ultrafine aerosol deposition in the human nasal and oral airways. J Aerosol Sci 27:785–801. https://doi.org/10.1016/0021-8502(96)00029-8

Crawford-Brown DJ, Shyr LJ (1987) The relationship between hit probability and dose for alpha emissions under selected geometries. Radiat Prot Dosimetry 20:155–168. https://doi.org/10.1093/oxfordjournals.rpd.a080024

Darby S, Hill D, Auvinen A, et al (2005) Radon in homes and risk of lung cancer: collaborative analysis of individual data from 13 European case-control studies. BMJ 330:223. https://doi.org/10.1136/bmj.38308.477650.63

Darby S, Hill D, others (2006) Residential radon and lung cancer—detailed results of a collaborative analysis of individual data on 7148 persons with lung cancer and 14 208 persons without lung cancer from 13 epidemiologic studies in Europe. Scand J Work Environ Health 32:1–84

Deak PD, Smal Y, Kalender WA (2010) Multisection CT Protocols: Sex- and Age-specific Conversion Factors Used to Determine Effective Dose from Dose-Length Product. Radiology 257:158–166. https://doi.org/10.1148/radiol.10100047

Farkas Á, Jókay Á, Balásházy I, et al (2016) Numerical simulation of emitted particle characteristics and airway deposition distribution of Symbicort® Turbuhaler® dry powder fixed combination aerosol drug. Eur J Pharm Sci 93:371–379. https://doi.org/10.1016/j.ejps.2016.08.036

Füri P, Farkas Á, Madas BG, et al (2020a) The degree of inhomogeneity of the absorbed cell nucleus doses in the bronchial region of the human respiratory tract. Radiat Environ Biophys 59:173–183. https://doi.org/10.1007/s00411-019-00814-0

Füri P, Groma V, Török S, et al (2020b) Ultrafine urban particle measurements in Budapest and their airway deposition distribution calculation. Inhal Toxicol 32:494–502. https://doi.org/10.1080/08958378.2020.1850937

Füri P, Hofmann W, Jókay Á, et al (2017) Comparison of airway deposition distributions of particles in healthy and diseased workers in an Egyptian industrial site. Inhal Toxicol 29:147–159. https://doi.org/10.1080/08958378.2017.1326990

Haefeli-Bleuer B, Weibel ER (1988) Morphometry of the human pulmonary acinus. Anat Rec 220:401–414. https://doi.org/10.1002/ar.1092200410

Haninger T (1998) Size distributions of radon progeny and their influence on lung dose. In: Radon and Thoron in the Human Environment. 7th Tohwa University International Symposium. World Scientific, Singapore, pp 574–576

Hofmann W (2011) Modelling inhaled particle deposition in the human lung—A review. J Aerosol Sci 42:693–724. https://doi.org/10.1016/j.jaerosci.2011.05.007

Hofmann W, Koblinger L (1990) Monte Carlo modeling of aerosol deposition in human lungs. Part II: Deposition fractions and their sensitivity to parameter variations. J Aerosol Sci 21:675–688. https://doi.org/10.1016/0021-8502(90)90122-E

Hofmann W, Koblinger L (1992) Monte Carlo modeling of aerosol deposition in human lungs. Part III: Comparison with experimental data. J Aerosol Sci 23:51–63. https://doi.org/10.1016/0021-8502(92)90317-O

Hofmann W, Sturm R (2004) Stochastic model of particle clearance in human bronchial airways. J Aerosol Med Off J Int Soc Aerosols Med 17:73–89. https://doi.org/10.1089/089426804322994488

International Commission on Radiation Units and Measurements (ICRU) (2012) Measurement and reporting of radon exposures. ICRU Report 88. J ICRU 12:1–191. https://doi.org/10.1093/jicru/ndv019

International Commission on Radiological Protection (ICRP) (1994) Human respiratory tract model for radiological protection. ICRP Publication 66. Ann ICRP 24:1–482

Koblinger L, Hofmann W (1990) Monte Carlo modeling of aerosol deposition in human lungs. Part I: Simulation of particle transport in a stochastic lung structure. J Aerosol Sci 21:661–674. https://doi.org/10.1016/0021-8502(90)90121-D

Kreuzer M, Auvinen A, Cardis E, et al (2018) Multidisciplinary European Low Dose Initiative (MELODI): strategic research agenda for low dose radiation risk research. Radiat Environ Biophys 57:5–15. https://doi.org/10.1007/s00411-017-0726-1

Madas BG, Füri P, Farkas Á, et al (2020) Deposition distribution of the new coronavirus (SARS-CoV-2) in the human airways upon exposure to cough-generated droplets and aerosol particles. Sci Rep 10:22430. https://doi.org/10.1038/s41598-020-79985-6

Marsh JW, Birchall A, Davis K (2005) Comparative dosimetry in homes and mines: estimation of K-factors. In: Radioactivity in the Environment. Elsevier, pp 290–298

Mercer RR, Russell ML, Crapo JD (1991) Radon dosimetry based on the depth distribution of nuclei in human and rat lungs. Health Phys 61:117–130

Mezey RJ, Cohn MA, Fernandez RJ, et al (1978) Mucociliary transport in allergic patients with antigen-induced bronchospasm[1–3]. Am Rev Respir Dis 118:677–684. https://doi.org/10.1164/arrd.1978.118.4.677

National Research Council (NRC) (1999) Health effects of exposure to radon: BEIR VI. National Academy Press (US), Washington, DC

Phalen RF, Oldham MJ (2001) Methods for modeling particle deposition as a function of age. Respir Physiol 128:119–130. https://doi.org/10.1016/S0034-5687(01)00270-5

Pleil JD, Ariel Geer Wallace M, Davis MD, Matty CM (2021) The physics of human breathing: flow, timing, volume, and pressure parameters for normal, on-demand, and ventilator respiration. J Breath Res 15:042002. https://doi.org/10.1088/1752-7163/ac2589

Raabe OG, Yeh HC, Schum GM, Phalen RF (1976) Tracheobronchial geometry: human, dog, rat, hamster. Lovelace Foundation, Albuquerque, New Mexico

Rogers D (2004) Airway mucus hypersecretion in asthma: an undervalued pathology? Curr Opin Pharmacol 4:241–250. https://doi.org/10.1016/j.coph.2004.01.011

Sturm R (2012) Theoretical models of carcinogenic particle deposition and clearance in children's lungs. J Thorac Dis 4:. https://doi.org/10.3978/j.issn.2072-1439.2012.08.03

United Nations Scientific Committee on the Effects of Atomic Radiation (UNSCEAR) (2000) Sources and effects of ionizing radiation: United Nations Scientific Committee on the Effects of Atomic Radiation: UNSCEAR 2000 report to the General Assembly, with scientific annexes. United Nations, New York

Yu CP, Diu CK, Soong TT (1981) Statistical analysis of aerosol deposition in nose and mouth. Am Ind Hyg Assoc J 42:726–733. https://doi.org/10.1080/15298668191420602